\documentclass[aps,prb,reprint,superscriptaddress,showpacs,amsmath,amssymb]{revtex4-1}

\usepackage{graphicx}
\usepackage{dcolumn}	
\usepackage{xspace}		
\usepackage{miller}		

\newcommand{\Tc}{\ensuremath{T_c}\xspace}
\newcommand{\Jc}{\ensuremath{J_c}\xspace}
\newcommand{\tmis}{\ensuremath{\theta_{\text{mis}}}\xspace}
\newcommand{\Bcr}{\ensuremath{B_{\text{cr}}}\xspace}

\newcommand{\REBCOformula}{REBa$_2$Cu$_3$O$_{7-\delta}$\xspace}

\newcommand{\unit}[1]{\ensuremath{~\mathrm{#1}}}
\newcommand{\unitnospace}[1]{\ensuremath{\mathrm{#1}}}
\newcommand{\K}{\unit{K}}
\newcommand{\T}{\unit{T}}

\newcommand{\A}{\unit{A}}
\newcommand{\nA}{\unit{nA}}
\newcommand{\pA}{\unit{pA}}
\newcommand{\kV}{\unit{kV}}
\newcommand{\muV}{\unit{\mu{}V}}
\newcommand{\mm}{\unit{mm}}
\newcommand{\mum}{\unit{\mu{}m}}
\newcommand{\nm}{\unit{nm}}
\newcommand{\degree}{\unitnospace{^\circ}}

\newcommand{\E}[1]{\ensuremath{\!\times\! 10 ^{#1}}}

\newcommand{\etal}{\textit{et~al.\ }}	
\newcommand{\vs}{vs\xspace}						
\newcommand{\ie}{i.e.\ }							

\begin{document}


\title{Individual grain boundary properties and overall performance of metal-organic-deposition coated conductors}


\author{M.~Weigand}
\affiliation{Department of Materials Science and Metallurgy, University of Cambridge, Pembroke St, Cambridge CB2 3QZ, United Kingdom}

\author{S.~C.~Speller}
\author{G.~M.~Hughes}
\affiliation{Department of Materials, University of Oxford, Parks Road, Oxford OX1 3PH, United Kingdom}

\author{N.~A.~Rutter}
\affiliation{Department of Materials Science and Metallurgy, University of Cambridge, Pembroke St, Cambridge CB2 3QZ, United Kingdom}

\author{S.~Lozano-Perez}
\author{C.~R.~M.~Grovenor}
\affiliation{Department of Materials, University of Oxford, Parks Road, Oxford OX1 3PH, United Kingdom}

\author{J.~H.~Durrell}
\email[]{jhd25@cam.ac.uk}
\affiliation{Department of Materials Science and Metallurgy, University of Cambridge, Pembroke St, Cambridge CB2 3QZ, United Kingdom}


\date{\today}

\begin{abstract}
We have investigated single grain boundaries (GBs) isolated in coated conductors produced by Metal-Organic Deposition (MOD). When a magnetic field is swept in the film plane, an angle-dependent crossover from boundary to grain limited critical current density $J_c$ is found. In the force-free orientation, even at fields as high as $8$~T, the GBs still limit $J_c$. We deduce that this effect is a direct consequence of GB meandering. We have employed these single GB results to explain the dependence of $J_c$ of polycrystalline tracks on their width: in-plane measurements become flatter as the tracks are narrowed down. This result is consistent with the stronger GB limitation at field configurations close to force-free found from the isolated boundaries. Our study shows that for certain geometries even at high fields the effect of GBs cannot be neglected.
\end{abstract}


\maketitle


\section{Introduction}

In high temperature superconductors it is often the presence of grain boundaries (GBs), rather than the inherent material properties, which limits the overall critical current density \Jc. Soon after their discovery it was found that \Jc decreases with increasing misorientation angle \tmis of the GB. \cite{Dimos88, Dimos90} Ivanov \etal \cite{Ivanov91} established that $\Jc(\tmis)$ follows an exponential dependence, which was subsequently confirmed by several studies. \cite{Verebelyi00, Feldmann01} For very low values of \tmis ($2$ -- $3\degree$) a plateau was observed, \ie \Jc of the boundary equals that of the surrounding grains. \cite{Feldmann01} The critical current density of GBs was also found, in general, to be less sensitive to applied magnetic fields than that of the grains. As a consequence above a crossover field \cite{Fernandez03} the overall \Jc is limited by the \emph{intra}granular value.

In the case of low angle boundaries with misorientation angles $\tmis < 10\degree$ the reduced value of \Jc was explained by dislocations at the boundaries because those defects reduce both the pinning strength and the cross sectional area for supercurrents. \cite{Chisholm91, Durrell09} As a direct consequence flux lines can channel along the boundary when they are aligned with the GB plane, \cite{Diaz98a, Hogg01} further reducing \Jc.

An alternative explanation for the suppression of \Jc at GBs is oxygen deficiency in the boundary region. \cite{Hilgenkamp02} In particular, this was found to be the case for high angle boundaries by TEM-EELS (electron energy loss spectroscopy). \cite{Babcock94} The same boundaries also showed significantly reduced critical current densities. Only minor deviations from the ideal oxygen content were observed in strongly coupled low angle GBs.

Due to its high critical current density the most promising material for practical conductors \cite{Larbalestier01} is \REBCOformula (REBCO), where RE is a rare earth, for example yttrium (YBCO). In order to achieve high transport currents a well textured superconducting layer is grown on an oriented substrate consisting of a metal tape buffered by oxide layers. This approach, termed coated conductors (CCs), ensures that all grain boundaries have only low angle misorientations (\ie $\tmis < 10\degree$), which causes the order parameter to be continuous across the boundaries. \cite{Redwing99} One of the most promising techniques to fabricate this complex structure is called RABiTS (Rolling Assisted Bi-axially Textured Substrates), \cite{Goyal96, Norton96} which is based on a textured Ni alloy tape.

The efforts to determine the electrical properties of single GBs can be divided into two groups. Firstly, there are experiments performed on superconductors grown on bicrystal substrates. These model systems, typically consisting of a single pure \hkl[001]-tilt boundary, allowed important insights to be gained into the mechanisms of current transport across geometrically simple GBs. \cite{Dimos90, Ivanov91, Diaz98, Diaz98a, Verebelyi00, Durrell03} Secondly, in more recent years grain boundaries have been isolated in actual coated conductor samples. \cite{Feldmann01, Feldmann07, Durrell07, Weigand09} The latter experiments are necessary because real CC boundaries usually show a combination of tilt and twist misorientation components \cite{Hilgenkamp02} which cannot easily be achieved in bicrystals. This is particularly interesting as there is currently an ongoing debate whether the in- or out-of-plane misalignment is more detrimental to current flow. \cite{Held09, Larbalestier09} Furthermore, certain \emph{ex situ} manufacturing routes lead to GBs which are not planar, like those found in films grown by PLD (pulsed laser deposition), but show a meandering morphology. \cite{Feldmann05, Feldmann06}

This meandering is due to the lateral growth mode of CC grains produced by physical vapor deposition and subsequent annealing of a BaF$_2$ based precursor \cite{Feenstra91} or by TFA-MOD (Metal-Organic Deposition using trifluoroacetates). \cite{Rupich03} Meandering, both along the length of the GB and through the film thickness, was found to enhance the critical current density of the boundaries without the need for complicated grain boundary doping (for example with Ca, Ref.~\onlinecite{Hammerl00}). As a consequence the exponential decay of \Jc with increasing \tmis is not followed. The beneficial properties of meandering GBs were explained by the combination of two mechanisms; firstly, the cross sectional area of the boundaries is increased, \cite{Feldmann07, Dinner07} and secondly vortex channeling is suppressed since vortices can lie in the plane of the GB only over short parts of their length. \cite{Durrell07}

A different approach to investigate the properties of grain boundaries is to measure wider (\ie polycrystalline) coated conductor tracks. Kim \etal \cite{Kim05} compared \Jc of films grown by MOD on a single crystal substrate and on RABiTS. They found that at $77\K$ \Jc of their (better) CC was suppressed with respect to the single crystal only up to fields of $\sim\!\!2\T$ applied perpendicular to the plane of the samples. Above this crossover field \cite{Fernandez03} grain boundaries do not limit the current flow any more. Kim \etal confirmed this result by successively narrowing down a CC track and measuring its $\Jc(B)$ dependence. At high fields \Jc was the same for all widths, whereas at low $B$ it decreased as the track got narrower. They concluded that this reduction was due to \Jc being limited by GBs at low fields. This was supported by the fact that the current-voltage curves obtained from their two narrowest tracks showed clear characteristics of GB dissipation. The crossover from boundary to grain limited \Jc was also observed in neutron irradiation experiments on CCs. \cite{Eisterer10} Irradiation reduces \Jc at low $B$ only, corroborating the fact that in this field regime the current is limited by GBs, which incur more damage by neutrons than grains.

The aim of the present study is to explore conditions when grain boundaries or grains limit \Jc of coated conductors at different magnitudes and orientations of applied magnetic fields and at different temperatures. The two approaches mentioned above were, therefore, combined. We first isolated different grain boundaries and a single grain in MOD samples. Their critical current densities were measured for magnetic fields applied in the plane of the film (Sec.~\ref{sec:single_GBs}). This type of measurement gives interesting insights into how microscopic currents flow, and it is also the predominant field orientation which occurs in several applications for coated conductors. We then extended the Kim \etal experiment \cite{Kim05} by investigating how \Jc depends on track width for in-plane fields. These results are presented in Sec.~\ref{sec:wider_tracks}. Not only do they deepen our understanding of current transport across GBs, but knowledge about the width dependence of \Jc is invaluable if striation is considered as a means of reducing ac losses. \cite{Glowacki01} In Sec.~\ref{sec:comparison} we finally discuss how the absolute \Jc values of the isolated grain and GBs relate to those of the polycrystalline tracks.


\section{Experimental}

\subsection{Samples analyzed}

The coated conductor samples analyzed in this study were grown on \mbox{RABiTS} using the American Superconductor \mbox{TFA-MOD} process. \cite{Rupich03, Rupich04, Schoop05} They had an architecture of YBCO ($\sim\!800\nm$) / CeO$_2$ ($\sim\!75\nm$) / YSZ (yttria-stabilized zirconia, $\sim\!75\nm$) / Y$_2$O$_3$ ($\sim\!75\nm$) / Ni-W ($\sim\!75\mum$). The YBCO grains varied in diameter from $20$ to $50\mum$, as determined by EBSD (Electron Backscatter Diffraction).


\subsection{Sample preparation}

\begin{figure}
	\includegraphics[width=\columnwidth]{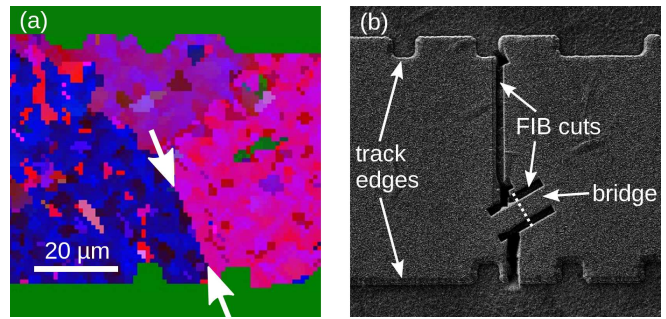}%
	\caption{\label{fig:EBSD_map_FIB_bridge}(Color online) (a) An EBSD map showing a grain boundary (marked by arrows) which was isolated subsequently. (b) FIB cuts were made to create a bridge in order to force the current across the GB in (a). The boundary position is indicated by white dots. Both images have the same scale.}
\end{figure}

In a first step, conventional photolithography and ion milling were used to pattern tracks $50\mum$ wide on samples cut from a CC tape. EBSD maps of areas consisting of several grains were then acquired, using a JEOL JSM6480LV microscope and HKL software. One of these maps is presented in Fig.~\ref{fig:EBSD_map_FIB_bridge}(a). They allowed us to select a grain and grain boundaries suitable for isolation and deduce the misorientation of neighboring grains bordering the GB.

Obtaining reliable orientation maps of YBCO films using EBSD is challenging, particularly after lithographic patterning. The surface topography, any residual surface contamination, and fine-scale mosaic structure of the films result in a poor pattern quality. \cite{Humphreys01, Koblischka03} In addition the Kikuchi patterns for $c$-axis and $a$-axis aligned YBCO are very similar, causing random misindexing by the software. The EBSD patterns have, therefore, been indexed using a cubic version of the unit cell.

In order to isolate the grain or GB of interest, bridges between $4.5$ and $5.0\mum$ wide were patterned within a single grain or spanning the GB, using a Zeiss Nvision~40 FIB/SEM system (Focused Ion Beam microscope), as shown in Fig.~\ref{fig:EBSD_map_FIB_bridge}(b). Ga contamination was kept to a minimum by carrying out all imaging with low ion beam currents ($\leq 10\pA$). This is confirmed by the fact that the transition temperature \Tc was reduced in patterned tracks by no more than $1\K$ compared to values obtained on unpatterned samples.

The successful isolation of grains and GBs in coated conductors has been reported in previous studies, \cite{Feldmann01, Feldmann07} using EBSD and conventional lithography. The advantage of our approach, first presented in Ref.~\onlinecite{Weigand09}, is that a FIB offers a significantly higher resolution when it comes to the positioning of the bridges. Consequently it allowed us to target specific grains and GBs more precisely. A FIB has also been used to isolate CC grain boundaries in a recent study on the effect of strain on \Jc of grains and GBs.\cite{Laan10}

In order to investigate how the critical current density depends on the width of polycrystalline tracks, we measured and successively narrowed down a $250\mum$ wide track, using photolithography and ion milling. Again no significant degradation of \Tc from the repeated patterning was found.


\subsection{Critical current density measurements}

Critical current densities were obtained by a four-terminal measurement at $65\K$ and $77.35\K$. Magnetic fields up to $8\T$ were applied perpendicular to the plane of the films and were swept in-plane using a two-axis goniometer. \cite{Herzog94} \Jc was determined using a voltage criterion of $0.5$ and $1.0\muV$ for the isolated grain/GBs and the wider tracks, respectively.


\subsection{TEM}

TEM lamella preparation was carried out using a Zeiss Nvision 40 FIB/SEM instrument. A layer consisting of electron beam deposited tungsten followed by ion beam deposited carbon was used to protect the surface of the specimen. Rough milling was carried out using Ga\textsuperscript{+} ion beam currents of $3.5\nA$ and then $750\pA$. The specimen was lifted out \textit{in situ} using a Kleindiek micromanipulator and mounted on a Cu grid using FIB-deposited C. Final FIB thinning was carried out with a beam current of $150\pA$. The SEM was constantly imaging during the final thinning process to monitor electron transparency and film thickness. TEM Bright Field (BF) imaging was performed in a Philips CM20 operated at $200\kV$. 


\section{Results and discussion}

\subsection{XRD}

The superconducting layer was well textured, as was established by a four circle X-ray diffractometer. The FWHM (full width at half maximum) of rocking curves obtained on the \hkl(005) peak was $3.2\degree$ and $4.4\degree$ along the rolling and transverse directions, respectively. A $\phi$-scan on the \hkl{103} peak confirmed the good in-plane alignment with a FWHM of $6.1\degree$.


\subsection{TEM}

\begin{figure}
	\includegraphics[width=\columnwidth]{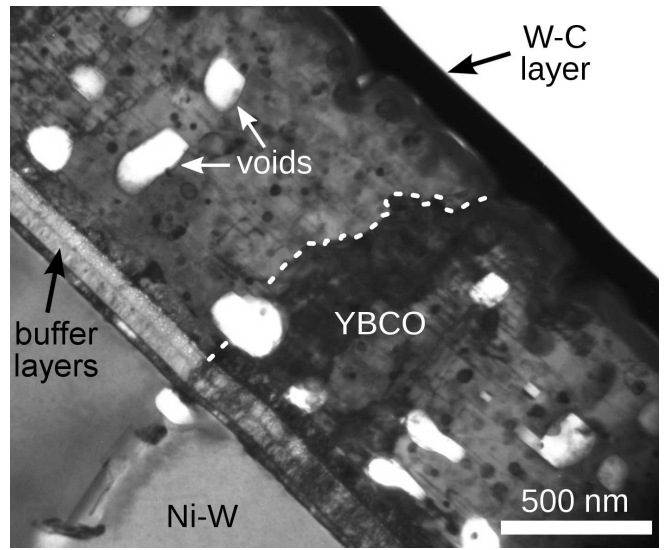}%
	\caption{\label{fig:TEM_cross-section}A cross-sectional TEM image of one of the boundaries which has been examined by transport measurements. \cite{Weigand09} The two grains can be distinguished from their different diffraction contrast and the GB is indicated with a dashed line. Many stacking faults and several voids are visible. The film surface (top right) is rough compared to the buffer layer--YBCO interface (bottom left).}
\end{figure}

The through-thickness meandering of boundaries in MOD films as reported previously \cite{Feldmann06, Feldmann07} was confirmed in our sample by a cross-sectional TEM image. The critical current density of the grain boundary depicted in Fig.~\ref{fig:TEM_cross-section} had been measured \cite{Weigand09} before a TEM sample was prepared from it. The two neighboring grains are clearly visible thanks to the diffraction contrast achieved by aligning a zone axis of one of the grains with the electron beam. Whereas the YBCO GB lines up with the substrate GB more closely than the boundary presented in Ref.~\onlinecite{Feldmann07}, its morphology is nevertheless very different from the planar boundaries characteristic of PLD films. The boundary meanders through the thickness of the film with amplitudes of less than $50\nm$. On the film surface the GB is displaced by about $300\nm$ from its location directly above the grain boundary in the buffer layer.

Stacking faults parallel to the $ab$-planes are present in the YBCO layer, separated from each other by $20$ to $50\nm$ along the $c$-axis. Several voids, between $50$ and $300\nm$ in diameter, are visible. Pores similar to these voids have been reported previously in TFA-MOD samples. \cite{Xia08, Matsuda08, Jang08} The surface of the film is relatively rough which leads to a variation in thickness at different positions of almost $100\nm$. The buffer layer--YBCO interface on the other hand is very smooth. We explain the asymmetry of in-plane \Jc measurements at low fields by this difference in roughness between the two opposed YBCO surfaces (see below).


\subsection{\label{sec:single_GBs}Isolated grain and grain boundaries}

Three individual grain boundaries (labeled GB1 -- GB3) and one single grain (IG, for \emph{intra}granular) were analyzed in this study and their key parameters are listed in Table~\ref{tbl:GB_samples}. EBSD maps also showed a certain amount of mosaicity within each grain, as in the upper grain in Fig.~\ref{fig:EBSD_map_FIB_bridge}(a) for instance. The misorientation angles \tmis are therefore averages of values obtained at several points along the length of each GB.

\begin{table}
	\caption{\label{tbl:GB_samples}Summary of the properties of the grain and GBs isolated for this study: crystallographic misorientation angles~\tmis, self-field \Jc, and crossover field \Bcr at maximum Lorentz force caused by a field in-plane and perpendicular to the bridge direction (both at $77.35\K$). The scatter in \tmis is due to grain mosaicity, rather than measurement error.}
	\begin{ruledtabular}
		\begin{tabular}{lrrd}
			Sample & \tmis (deg) & $\Jc(\text{sf})$ (\unitnospace{A\,m^{-2}}) & \Bcr \text{ (T)} \\
			\hline
			IG & (grain) & $6.08\E{10}$ & - \\
			GB1 & $4.9\pm0.4$ & $4.98\E{10}$ & <0.25 \\
			GB2 & $5.7\pm0.9$ & $4.36\E{10}$ & 0.75 \\
			GB3 & $6.5\pm1.3$ & $2.17\E{10}$ & 3 \\		
		\end{tabular}
	\end{ruledtabular}
\end{table}

\begin{figure}
	\includegraphics[width=\columnwidth]{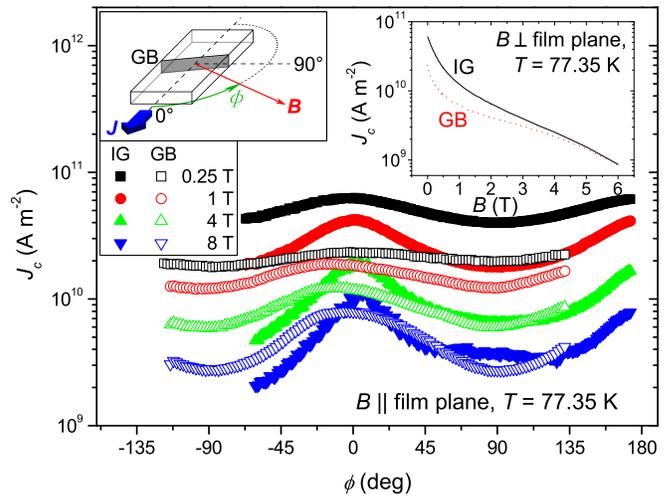}%
	\caption{\label{fig:phiscans_B14IG_B29GBB}(Color online) In-plane scans for the isolated grain and one of the grain boundaries (GB3). It can be seen clearly that at higher fields the \Jc of grain and GB overlap at orientations around maximum Lorentz force ($\phi \approx 90\degree$) whereas at angles around the force free orientation ($\phi \approx 0\degree$) the grain has superior properties for all fields. The inset on the left shows the measurement geometry for in-plane scans: the FIB bridge is sketched together with the GB it crosses. $\Jc(B)$ obtained on the same grain and GB for fields perpendicular to the film surface is depicted in the second inset. As for fields in-plane, a crossover from GB to grain limited \Jc is found.}
\end{figure}

Figure~\ref{fig:phiscans_B14IG_B29GBB} shows \Jc of the isolated grain and one grain boundary (GB3) at $77.35\K$ for magnetic fields swept in the plane of the sample: $\phi = 0\degree$ corresponds to the force free (FF) orientation, while at $\pm90\degree$ the (macroscopic) current direction leads to maximum Lorentz force (see left inset of Fig.~\ref{fig:phiscans_B14IG_B29GBB}). The most striking feature is that at low fields \Jc of the GB is suppressed significantly with respect to the grain, whereas at high fields they roughly overlap. This behavior is consistent with the crossover from GB to grain limited critical current density for $B \perp$ film plane reported previously in several studies. \cite{Fernandez03, Kim05, Hanisch05, Feldmann07}

Upon closer inspection, however, it can be seen that at angles around the force free orientation ($\phi \approx 0\degree$) even at $8\T$ the grain still has a somewhat higher \Jc than the GB. Around maximum Lorentz force orientations, on the other hand, they overlap above a crossover field \Bcr. This is the opposite behavior of what was found by Durrell \etal \cite{Durrell03} for grain boundaries in films grown by PLD on bicrystal substrates. They showed that \Jc of the boundary was only reduced compared to a single grain when the field was within a certain angle $\phi_k$ of the GB plane (which corresponds to $\phi \approx 90\degree$ for our measurement geometry).

\begin{figure}
	\includegraphics[width=\columnwidth]{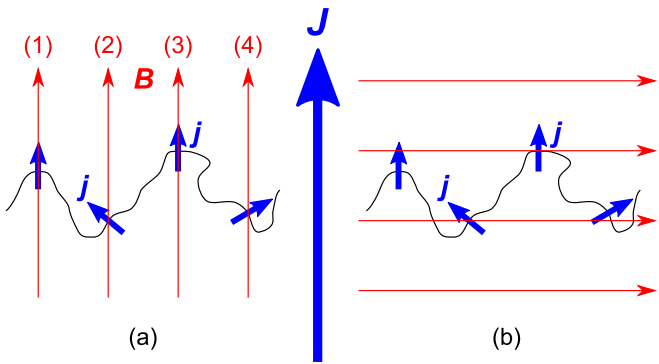}%
	\caption{\label{fig:currents_across_meandering_GB}(Color online) Current flow across a meandering grain boundary for (a) the macroscopic force free orientation ($\phi=0\degree$) and (b) maximum Lorentz force ($\phi=90\degree$). Due to the meandering a significant proportion of the microscopic currents $j$ are not parallel or perpendicular to the vortices. Consequently, some flux lines do not experience minimum or maximum Lorentz force, respectively, as would be the case if all currents were parallel to the macroscopic current direction $J$. For instance, in (a) vortices (2) and (4) are exposed to a Lorentz force $>0$. This is the reason why $\Jc(\phi=0\degree)$ is suppressed, whereas at $90\degree$ we find an improved behavior of the meandering GB, compared to a planar GB.}
\end{figure}

The key to understanding this behavior is found in the characteristic grain boundary meandering almost always observed in samples prepared by chemical reaction routes,\cite{Feldmann05, Feldmann06, Laan10} and seen in our samples both in the tape plane and through the film thickness [see Fig.~\ref{fig:EBSD_map_FIB_bridge}(a) and \ref{fig:TEM_cross-section}]. In a field and temperature regime where \Jc is limited by the GB, microscopic currents cross it perpendicular to the specific boundary segments. \cite{Dinner07} They flow in many different directions, as is illustrated in Fig.~\ref{fig:currents_across_meandering_GB}(a). Consequently at $\phi = 0\degree$ a significant number of the vortices do not experience zero Lorentz force as they do in a single grain or a planar PLD grain boundary. In the latter two cases all currents flow parallel to the macroscopic current direction or in one direction perpendicular to the GB, respectively. Because of the different way in which currents cross an MOD GB we would expect it always to have a reduced \Jc at the macroscopic force free orientation.

At $\phi = \pm90\degree$ meandering improves the performance of a grain boundary. Whereas all vortices in the grain are subject to maximum Lorentz force, a large proportion of the flux lines in the GB experience a smaller force since microscopic currents flow at angles $<90\degree$ relative to them [see Fig.~\ref{fig:currents_across_meandering_GB}(b)]. We therefore find \Jc of the grain and the GB to overlap above a certain field, which means that in the case of the bridge across the boundary it is in fact the grains on either side, rather than the GB, which limit \Jc.

The right inset of Fig.~\ref{fig:phiscans_B14IG_B29GBB} shows $\Jc(B)$ for fields applied perpendicular to the film plane. As reported previously \cite{Feldmann07} we also find a crossover from GB to grain limited critical current density at this field configuration: at $B \approx 5\T$ \Jc of grain and GB become equal. This field is notably higher, however, than that reported in Ref.~\onlinecite{Feldmann07}.

\begin{figure}
	\includegraphics[width=\columnwidth]{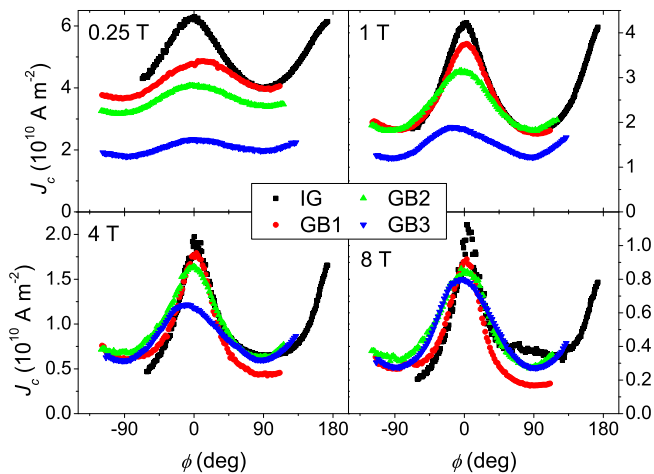}%
	\caption{\label{fig:phiscans_all_GBs}(Color online) The critical current density of all three grain boundaries and the single grain at $77.35\K$ for magnetic fields of $0.25$, $1$, $4$, and $8\T$ swept in the plane of the films. For the least misaligned boundary (GB1) the crossover from GB to grain dominated behavior occurs first and is present over the largest angular range as the field increases.}
\end{figure}

Figure~\ref{fig:phiscans_all_GBs} presents in-plane scans of the grain and the GB discussed above together with data obtained on two other boundaries. It can be seen in the $0.25\T$ scans that as expected at low fields \Jc decreases monotonically with increasing misorientation angle (see also Table~\ref{tbl:GB_samples}).

Grain boundary GB1 has the lowest crystallographic misorientation of the three GBs analyzed. This is why even at a field as small as $0.25\T$ its \Jc equals that of the grain at $\phi$ around $90\degree$. This region broadens as we increase the field (see Fig.~\ref{fig:phiscans_all_GBs}) until at $8\T$ only at angles close to the force free orientation does the grain still show a slightly higher critical current density.

At $B<\Bcr$ the critical current density of GB2 lies between the curves obtained on GB1 and GB3 at all angles, consistent with its intermediate misorientation angle. The crossover at maximum Lorentz force occurs at a field of $0.75\T$ confirming that \Bcr increases monotonically with \tmis (see Table~\ref{tbl:GB_samples}).

The data from a fourth boundary (presented in Ref.~\onlinecite{Weigand09}) showed a completely different behavior. At low fields its critical current density was virtually independent of the in-plane angle $\phi$. Only from $B = 1\T$ upwards was a dip found in $\Jc(\phi)$ at about $90\degree$. Around this angle \Jc of the GB overlapped almost exactly with that of the single grain, which also showed a minimum at $\phi=90\degree$. This can again be explained by a current limitation caused by the grains adjacent to the boundary. At present it is not clear why the GB in Ref.~\onlinecite{Weigand09} behaved in a different way from the boundaries presented here. A likely explanation would be a \Jc limitation in a direction perpendicular to the plane of the sample that was not detected by EBSD imaging on the surface, causing \Jc to be independent of $\phi$.

\begin{figure}
\includegraphics[width=\columnwidth]{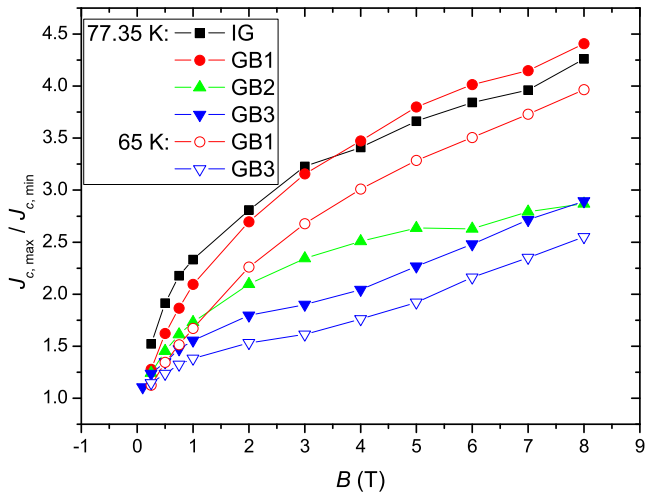}%
	\caption{\label{fig:in-plane_anisotropy_B14IG_B29GBs_B21GB}(Color online) The in-plane anisotropy decreased with increasing misorientation angle and with decreasing temperature.}
\end{figure}

The difference in \Jc at FF and maximum Lorentz force orientations can be described by the in-plane anisotropy $\zeta = \Jc($FF$) / \Jc($max.\ LF$)$, which is plotted \vs field in Fig.~\ref{fig:in-plane_anisotropy_B14IG_B29GBs_B21GB} for GB1--3 and the single grain (IG). As expected at low fields, $\zeta$ is highest for the single grain. Unlike in the bridges containing a GB, microscopic currents flow parallel to the macroscopic current direction, leading to a strong FF maximum and a high $\zeta$ value. Above $2\T$, where \Jc of the single grain and GB1 overlap over almost the entire angular range, their in-plane anisotropies are also very similar. The lower values of the critical current densities of GB2 and GB3 at FF on the other hand are reflected in a lower value of $\zeta$.

$\Jc(\phi)$ of the single grain and GB1 were asymmetric at fields $B > 1\T$, \ie the minima at $\phi = +90\degree$ and $-90\degree$ were found at different values (see Fig.~\ref{fig:phiscans_all_GBs}, data for $4$ and $8\T$). We explain this by a certain amount of out-of-plane tilt of the limiting grain(s). In order to align the sample parallel to the applied magnetic field it was tilted about the second goniometer axis, described by the angle $\theta$ between the direction of applied field and the sample normal. $\theta$-scans had been performed at $\phi = +90\degree$ (single grain) and $\phi = -90\degree$ (GB1), the maximum of which gave the value of $\theta$ where $B||ab$. If the sample is now rotated in $\phi$ by $180\degree$, however, the $ab$-planes are not parallel to the field any more, due to the misalignment between film surface and grain(s). As a consequence the second minimum is suppressed with respect to the first one.

\begin{figure}
	\includegraphics[width=0.85\columnwidth]{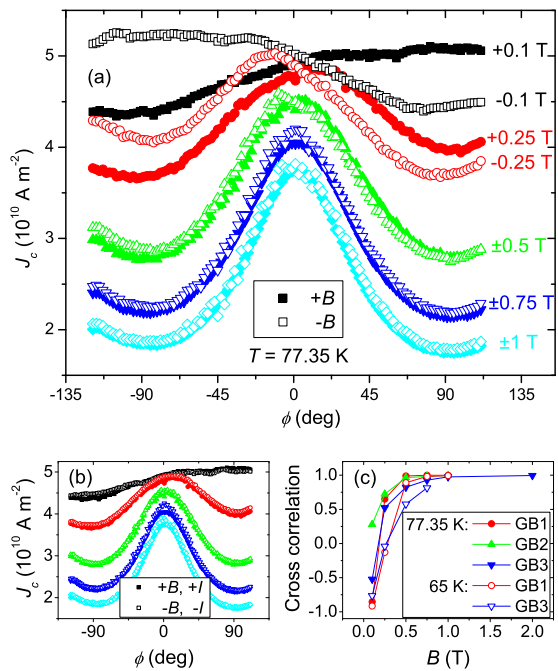}%
	\caption{\label{fig:phiscans_B29GBD_asymmetry}(Color online) (a) At low fields the in-plane scans became asymmetric, as is shown here for boundary GB1. Changing sign of the applied field reverses this behavior. The asymmetry can be explained by different barriers for flux entry from the film surface and through the substrate. (b) By reversing field \emph{and} current the original result is recovered: ($+B$, $+I$) equals ($-B$, $-I$). The colors represent the same fields as in (a). (c) The cross correlation for positive and negative fields shows that asymmetry persists to higher fields the higher the GB misorientation and the lower the temperature.}
\end{figure}

At low fields up to about $0.5\T$, $\Jc(\phi)$ of all three grain boundaries showed asymmetric behavior of a different kind. As can be seen in Fig.~\ref{fig:phiscans_B29GBD_asymmetry}(a), the asymmetry is reversed upon change of sign of the applied magnetic field. We therefore conclude that it is caused by a different surface barrier \cite{Bean64a} for flux entry through the film surface and the substrate--YBCO interface. Positive $\phi$ and positive fields correspond to flux entry through the substrate, which gives a higher \Jc than when the vortices enter from the top of the samples. The surface of the films is rough compared to the substrate--YBCO interface, as was shown by TEM. We reason that this roughness causes increased localized fields due to the demagnetization effect, making it easier for flux lines to penetrate the sample in this direction and hence suppress \Jc. \cite{Harrington09} This explanation is supported by the fact that the asymmetry is also reversed when the current direction is changed --- Fig.~\ref{fig:phiscans_B29GBD_asymmetry}(b) shows that positive applied fields and positive currents give the same results as negative $B$ and $I$. This also applies to the inverse configurations, \ie ($+B$, $-I$) gives the same results as ($-B$, $+I$) (not shown).

In order to quantify the asymmetry the cross correlation
\begin{equation}
r = \frac{\sum_i [(x_i - \overline{x}) (y_i - \overline{y})]}{\sqrt{\sum_i (x_i - \overline{x})^2} \sqrt{\sum_i (y_i - \overline{y})^2}}
\label{eq:cross_correlation}
\end{equation}
was determined for each set of $\phi$-scans at a given positive and negative applied field ($x_i$ is the value of \Jc at a specific $\phi_i$ for a certain positive $B$, $y_i$ for the same $\phi_i$ and negative $B$, and $\overline{x}$ and $\overline{y}$ are the averages of the entire $\phi$-scans). Figure~\ref{fig:phiscans_B29GBD_asymmetry}(c) shows that the field where $r$ becomes $1$, corresponding to a perfect overlap of the curves for positive and negative field and thus the disappearance of asymmetry, shifts to higher fields as the grain boundary misorientation angle becomes higher (see Table~\ref{tbl:GB_samples}). We thus conclude that a stronger GB limitation enhances asymmetry. This is consistent with the fact that asymmetry persists up to higher fields at $65\K$ (where more influence of GBs is expected, see below) than at $77.35\K$.

As can be seen in Fig.~\ref{fig:phiscans_B29GBD_asymmetry}(a), \Jc at the FF orientation was almost the same at $0.1$ and $0.25\T$. Weak field dependence is in general associated with critical current density governed by GBs. \cite{Durrell03, Durrell07} We therefore conclude that \Jc is strongly GB limited in this field and angular range, which is consistent with the observation that the crossover to grain limited behavior occurs first at maximum Lorentz force.

The fact that at very low fields \Jc of the bridges with a grain boundary is completely limited by the GB could also explain the peculiar behavior at $0.1\T$ (see Fig.~\ref{fig:phiscans_B29GBD_asymmetry}(a)). At this field the critical current density exhibits a broad maximum at $\phi \approx +90\degree$ and a minimum at $-90\degree$. At the macroscopic force free configuration we measure an intermediate value. We reason that due to the strong meandering at low fields the direction of microscopic currents covers at least the angular range of $\pm90\degree$ relative to the macroscopic current direction. If all directions occur with the same frequency we would expect no angular dependence of \Jc of the bridge at all. At $\phi=-90\degree$, however, all (or at least the vast majority of) vortices experience a Lorentz force pointing from the sample surface to the substrate, \ie along the direction with a lower surface barrier. At $+90\degree$ the opposite is the case. At $0\degree$ half the vortices are pushed in one direction and the other half in the other, which explains why $\Jc(-90\degree) < \Jc(0\degree) < \Jc(+90\degree)$.

The boundaries GB1 and GB3 were also measured at $65\K$ (not shown), and qualitatively similar behavior was found as at $77.35\K$. Again the $\phi$-scans were asymmetric for both GBs at low fields and for GB1 also at high fields. As expected at low fields \Jc of the $6.5\degree$ boundary was suppressed with respect to the $4.9\degree$ boundary at all angles. At maximum Lorentz force both boundaries could support the same current from $\sim\!4\T$ upwards, whereas no overlap was found for the force free orientation, where even at $B = 8\T$ \Jc of GB1 surpassed that of GB3 by a factor of $1.7$. An increased crossover field at a lower temperature is consistent with results for $B \perp$ film plane. \cite{Fernandez03} As was shown above, strong GB limitation at FF causes a flattening of $\phi$-scans, and in fact the in-plane anisotropy was reduced at the lower temperature, as can be seen in Fig.~\ref{fig:in-plane_anisotropy_B14IG_B29GBs_B21GB}. This behavior can also be understood in terms of increased current percolation which is expected when a CC becomes more GB dominated and which leads to flatter $\phi$-scans. \cite{Rutter05} It fits well into this picture that at $65\K$ $\Jc(\phi = 0\degree)$ is almost completely independent of field up to $B = 0.5\T$, which is higher than that found for $77.35\K$.


\subsection{\label{sec:wider_tracks}Tracks several grains wide}

\begin{figure}
	\includegraphics[width=\columnwidth]{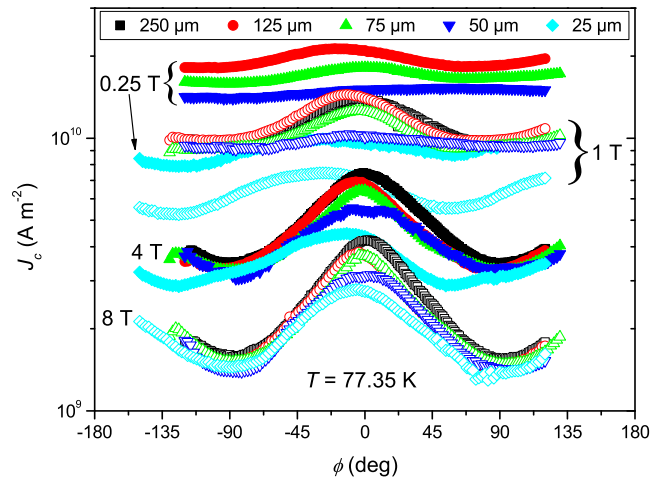}%
	\caption{\label{fig:phiscans_different_widths}(Color online) In-plane measurements of a track which has been narrowed down from $250$ to $25\mum$. At low fields, like $0.25\T$, \Jc decreases with decreasing track width over the whole angular range while at higher fields it is only suppressed at angles around the force free orientation ($\phi \approx 0\degree$). The measurement geometry is the same as for the isolated GBs (see inset of Fig.~\ref{fig:phiscans_B14IG_B29GBB}).}
\end{figure}

The critical current density for different magnetic fields swept in the plane of a track which has been successively reduced in width and re-measured can be seen in Fig.~\ref{fig:phiscans_different_widths}. At $0.25\T$ \Jc decreases for all angles as the track becomes narrower. (Note that low field data for the $250\mum$ track is missing because of our $5\A$ current limitation.) This behavior is consistent with the results obtained by Kim \etal \cite{Kim05} for fields applied perpendicular to the plane of the film. They explained their findings by the critical current density being limited by the GB network at low fields, rather than by the grains. A boundary segment with a low critical current density has a stronger effect on a narrow track, compared to a wider one where currents can percolate around the weak segment. A reduction in width can thus be expected to lead to a suppression in overall \Jc, as long as track length $\gg$ grain diameter, and therefore it is very likely that there is a weak GB in every track. This condition should be fulfilled in our experiment with the tracks being $1\mm$ long.

At $0.75\T$ the curves we obtained on the $75$ and the $50\mum$ track overlap at around $\phi = \pm90\degree$. In comparison the $125\mum$ track still shows a higher, the $25\mum$ track a lower, \Jc at all angles. As we increase the field further \Jc at maximum Lorentz force becomes approximately the same for all track widths (see Fig.~\ref{fig:phiscans_different_widths}), again a behavior reported previously for $B \perp$ film plane. \cite{Kim05} (In order to check for consistency we also performed measurements for fields perpendicular to the sample plane (not shown), which reproduced the results in Ref.~\onlinecite{Kim05}.) At the force free orientation on the other hand the wider tracks still exhibit a superior performance; even at $8\T$ the critical current density for angles around $\phi = 0\degree$ is reduced in the same sequence as is the track width.

\begin{figure}
	\includegraphics[width=\columnwidth]{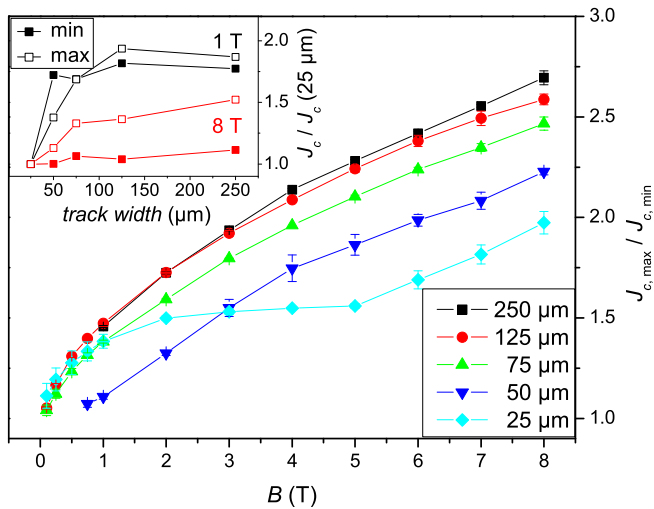}%
	\caption{\label{fig:in-plane_anisotropy_different_widths}(Color online) The in-plane anisotropy $\zeta = \Jc($FF$) / \Jc($max. LF$)$ becomes lower with decreasing track width above $1\T$. Some $\phi$-scans were slightly asymmetric (\ie the values of the minima at $+90\degree$ and $-90\degree$ differed), hence the error bars in this graph. The inset shows that at high fields, like $8\T$, the minima of the $\phi$-scans become independent of track width, whereas the maxima decrease as the track is narrowed down.}
\end{figure}

In order to quantify the dependence of \Jc on track width, the values of the maxima and the averages of the two minima are plotted \vs track width in the inset of Fig.~\ref{fig:in-plane_anisotropy_different_widths}. The data were scaled to the corresponding values of the $25\mum$ track. It can be seen that at $1\T$ both minimum and maximum depend on width. The critical current density of the maximum decreases with decreasing track width also at $8\T$, whereas \Jc of the minimum remains constant at this field.

We can explain these findings using the results on single grain and GBs presented above. At $\phi \approx \pm90\degree$ even at relatively low fields the grains, rather than the boundaries, limit the current carrying capability of the bridges with the isolated GBs. Due to the good out-of-plane alignment of the grains, as shown by XRD, we expect all grains to have a very similar \Jc, leading to a high level of homogeneity across the width of the track. As a consequence, narrowing a track down has no noticeable effect on its \Jc in this regime. For $\phi \approx 0\degree$ on the other hand we have demonstrated that GBs can carry only a reduced current compared to grains, even at high applied fields. In a wide track a boundary segment with a high misalignment does not cause a significant reduction of the overall \Jc as long as the adjacent boundaries have better properties, allowing current to percolate around the inferior GB segment. Apparently this is still the case for a track $125\mum$ ($3$--$6$ grains) wide, as we find a very similar behavior for the $250$ and the $125\mum$ tracks. Once the track is narrowed down to three or fewer grains in width, however, the effect of a weak boundary is not negligible any more and we measure a depressed \Jc at the force free orientation.

The qualitatively different behavior for varying track width becomes even more apparent when we plot the in-plane anisotropy $\zeta = \Jc($FF$) / \Jc($max.\ LF$)$ \vs applied field. As can be seen in Fig.~\ref{fig:in-plane_anisotropy_different_widths} this value is almost exactly the same for the two widest tracks. In the $75$, $50$ and $25\mum$ tracks, however, $\zeta$ is reduced, which corresponds to the flatter $\phi$-scans.

The increase of $\zeta$ with field can be understood as follows. At low $B$ the critical current density is predominantly GB dominated and the microscopic currents percolate strongly, which leads to a suppression of the force free effect and thus a low in-plane anisotropy. \cite{Rutter05} As we increase the field GBs stop being a barrier for current flow over large angular ranges, as was shown above, and the tape behaves more like a single crystal. Consequently microscopic currents flow more closely parallel to the macroscopic current direction, causing a strong force free maximum and a high value of $\zeta$.

It is worth pointing out that $\zeta$ is virtually independent of track width at $B \leq 1\T$ (with the exception of the $50\mum$ track, see below). Only for higher fields do the curves branch out and we find a monotonic decrease of the in-plane anisotropy as the width is reduced. This is particularly interesting as \Jc, both for $B \perp$ film plane (Ref.~\onlinecite{Kim05}) and for $B$ in-plane at maximum Lorentz force (present study), was found to be independent of track width at \emph{high} fields while at low fields narrower tracks showed lower values. The fact that $\zeta$ follows the opposite trend is clearly due to \Jc depending on width in a very different way for minimum and maximum Lorentz force. Consequently even at $8\T$ the effect of grain boundaries is still detectable.

\begin{figure}
	\includegraphics[width=\columnwidth]{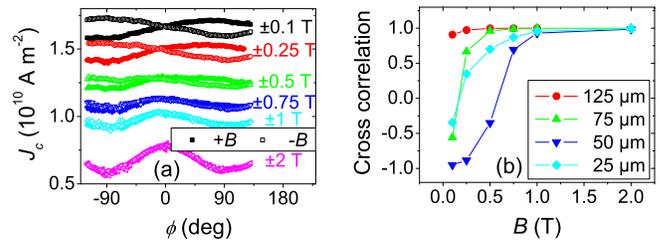}%
	\caption{\label{fig:phiscans_50_micron_asymmetry}(Color online) (a) As the track was narrowed down, in-plane scans showed asymmetric behavior at low fields, as is shown here for a track $50\mum$ wide. (b) The cross correlation between curves obtained at the same field but with opposite sign allows a quantification of this asymmetry. In general asymmetric behavior persists up to higher fields the narrower the track is.}
\end{figure}

The asymmetry of the $\phi$-scans at low fields found for single grain boundaries was also present in the curves obtained on the polycrystalline tracks. While not very significant in the $250$ and $125\mum$ wide links, this phenomenon became very pronounced after the track had been narrowed down further, as is depicted in Fig.~\ref{fig:phiscans_50_micron_asymmetry}(a) for a width of $50\mum$. Again changing the sign of the applied field or current polarity reversed this behavior. The cross correlation $r$ was determined according to Eq.~(\ref{eq:cross_correlation}) for each track width. Figure~\ref{fig:phiscans_50_micron_asymmetry}(b) shows that the field where $r$ becomes $1$ (disappearance of asymmetry) shifts to higher fields as the track gets narrower. For single GBs we found that asymmetry persists up to higher fields for higher misorientation angles. It is thus consistent that this is also the case for narrower tracks which are expected to be more dominated by boundaries than wider tracks. The exception is again the $50\mum$ track: $r$ almost exactly equals $-1$ at $B=0.1\T$ (maximum but inverse correlation) and a field higher than for all other tracks needs to be applied in order to recover symmetric behavior.

At low fields the $50\mum$ track showed only one minimum and one maximum at $\phi \approx \pm90\degree$, respectively [see Fig.~\ref{fig:phiscans_50_micron_asymmetry}(a)], similar to what was observed for GB1 [see Fig.~\ref{fig:phiscans_B29GBD_asymmetry}(a)]. As this track is only one or two grains wide it is reasonable to assume that the same explanation applies, \ie that minimum and maximum are related to a difference in the surface barrier.

Both minima and the maximum of \Jc of the $25\mum$ wide track shifted by $\sim\!30\degree$ over the field range investigated (see Fig.~\ref{fig:phiscans_different_widths}). At $B = 8\T$ they are found at the angles where maximum or minimum Lorentz force, respectively, occur for macroscopic current direction. As the field decreases they move to lower values of $\phi$. We reason that this is due to the crossover from grain to GB limited \Jc and presume that the limiting GB is not exactly perpendicular to the track. This would lead to an average direction of current flow not parallel to the track, causing the shift of maximum Lorentz force and force free orientations. This behavior is independent of the sign of $B$ and therefore must not be confused with the asymmetry discussed above. Further evidence that the $25\mum$ track is the one that is most GB dominated can be found in the fact that in this track the $0.1$ and $0.25\T$ $\phi$-scans are closer together than in the wider ones.

The surprisingly flat in-plane scans of the $50\mum$ wide track at lower fields might also be explained by a GB in the $25\mum$ track which is not perpendicular to the track direction. One could speculate that the rest of the GB segments making up the limiting path of the $50\mum$ track are at a significantly different angle than those in the $25\mum$ track. It would then follow that the maximum and minima get smeared out because at no particular value of $\phi$ is the majority of the limiting path perpendicular or parallel to the direction of applied field. In order to prove such a ``macroscopic meandering'' effect, however, further investigations would be necessary.


\subsection{\label{sec:comparison}Comparison between isolated grain/GBs and wider tracks}

\begin{figure}
	\includegraphics[width=\columnwidth]{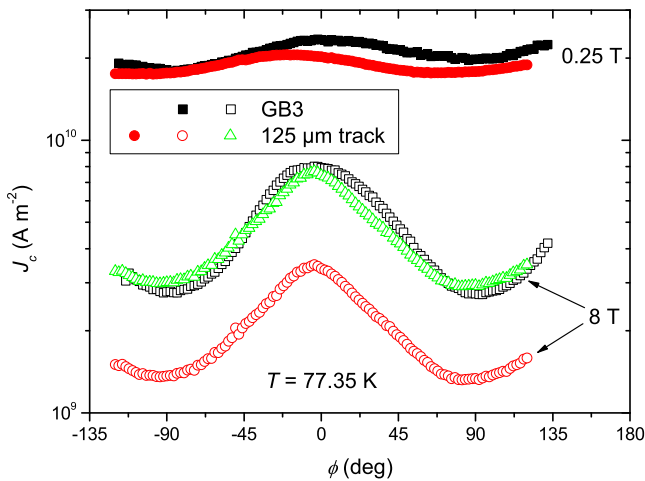}%
	\caption{\label{fig:comparison_B29GBB_125micron}(Color online) At $0.25\T$ the in-plane scan of GB3 is very similar to that of the $125\mum$ wide track. When the field is increased to $8\T$, $\Jc(\phi)$ obtained on the polycrystalline track ($\bigcirc$) is reduced compared to GB3 ($\square$) if a $0.5\muV$ criterion is used. If the electrical field criterion of the $125\mum$ track is changed ($\triangle$) so it matches that of the boundary, they overlap.}
\end{figure}

It is worth comparing the absolute \Jc values of the in-plane scans of isolated grain and GBs to those of the wider tracks. As a representative example curves obtained on the $125\mum$ wide track are plotted in Fig.~\ref{fig:comparison_B29GBB_125micron}. At $0.25\T$ the critical current density of the polycrystalline track approximates that of GB3 (note that $\Jc(\phi)$ of the $125\mum$ track has been re-analyzed using a $0.5\muV$ criterion, so it matches that used for GB3). It is  a reasonable assumption that at low fields the limiting path is defined by one set of boundaries crossing the track whose average \Jc equals that of GB3.

As we increase the field we find that \Jc of the $125\mum$ track is significantly below that of GB3 (see the $8\T$ curves in Fig.~\ref{fig:comparison_B29GBB_125micron}). The explanation for this can be found in the different length of the FIB bridge ($\sim\!\!15\mum$) and the polycrystalline track ($1000\mum$). As we can expect \Jc to be limited by the properties of the grains over almost the entire angular range at high $B$, the voltage drop occurs over the whole length of the bridge or track when \Jc is exceeded. It is, therefore, not valid to use the same \emph{voltage} criterion for both the FIB bridge and the wide track. Instead we need to apply the same \emph{electrical field} criterion, which implies a voltage criterion for the polycrystalline track $1000/15$ times as large as that for the FIB bridge. During our measurements current-voltage curves ($IV$s) were obtained up to $\sim\!\!7\muV$ only (in order to minimize the risk of damaging the track). Linear extrapolation of $IV$s at different $\phi$ shows that \Jc of the $125\mum$ track obtained using the correct $E$-field criterion is $\sim\!\!2.2$ times as high as that deduced from a $0.5\muV$ criterion. We therefore multiplied the $8\T$ scan of the $125\mum$ track by this value, and it can be seen in Fig.~\ref{fig:comparison_B29GBB_125micron} that it now overlaps almost perfectly with $\Jc(\phi)$ of GB3.


\section{Conclusions and summary}

We have isolated single grain boundaries (GBs) and a grain in an MOD coated conductor and measured their critical current density for magnetic fields swept in the film plane. In a second experiment we have investigated the dependence of \Jc on the width of a polycrystalline track, again for in-plane fields. 

In both cases two regimes, depending on both in-plane angle and magnitude of field, can be distinguished: (1) low $B$ regardless of angle, as well as high $B$ around the force free (FF) orientation, and (2) elevated fields around the maximum Lorentz force configuration. In regime (1) the isolated GBs show a suppressed \Jc with respect to the grain; in (2) \Jc of all GBs and the single grain become the same. This means that boundaries do not limit the current flow any more in the latter case. The angle dependent crossover from GB to grain limited behavior can be explained by the fact that MOD boundaries are not planar but meander. Microscopic currents do not, in general, flow parallel to the macroscopic current direction. As a consequence GBs behave comparably better at macroscopic maximum Lorentz force than at the force free orientation. It is remarkable that, despite the meandering, grain boundary limitation at FF persists up to $8\T$, the highest field analyzed.

In the case of the polycrystalline track it was found that in regime (1) \Jc decreases with decreasing track width. In (2), on the other hand, all track widths give the same \Jc. This is consistent with data from the isolated GBs, which only in regime (1) have an inferior \Jc compared to the grain. A GB limitation of the critical current density is, therefore, responsible for its width dependence, which persists up to and beyond fields of $8\T$. This behavior leads to the interesting result that the in-plane anisotropy depends on track width at high but not at low fields. We conclude that in applications with a strong in-plane component of the field the effect of GBs must be taken into account even at high $B$.


\begin{acknowledgments}
One of us (M.~W.) would like to thank M.~E.~Vickers for help with XRD measurements. We are grateful to American Superconductor for supplying the samples. This work was supported by the UK Engineering and Physical Sciences Research Council. The use of an SEM for EBSD analysis was made possible through the Oxford University Department of Materials BegbrokeNano facilities funded by the UK Government's Micro/Nano Technology Programme.
\end{acknowledgments}


%

\end{document}